\documentclass[pra,twocolumn,superscriptaddress]{revtex4-1}

\usepackage[utf8]{inputenc}
\usepackage[english]{babel}
\usepackage[T1]{fontenc}
\usepackage{lmodern}

\usepackage{graphicx}
\usepackage{amsmath}
\usepackage{amsfonts}
\usepackage{amssymb}
\usepackage{amsthm}

\usepackage{mathtools}
\usepackage{physics}

\usepackage{stmaryrd}

\usepackage{dsfont}
\usepackage{microtype}

\usepackage[breaklinks=true,colorlinks]{hyperref}

\newcommand{\cf}{cf.\ }

\begin{document}

\title{Polynomial scaling enhancement in ground-state preparation of Ising spin models via counter-diabatic driving}

\author{Andreas Hartmann}
\email{andreas.hartmann@uibk.ac.at}
\affiliation{Institut f\"ur Theoretische Physik, Universit\"at Innsbruck, Technikerstra{\ss}e~21a, A-6020~Innsbruck, Austria}
\author{Glen Bigan Mbeng}
\email{glen.mbeng@uibk.ac.at}
\affiliation{Institut f\"ur Theoretische Physik, Universit\"at Innsbruck, Technikerstra{\ss}e~21a, A-6020~Innsbruck, Austria}
\author{Wolfgang Lechner}
\email{wolfgang.lechner@uibk.ac.at}
\affiliation{Institut f\"ur Theoretische Physik, Universit\"at Innsbruck, Technikerstra{\ss}e~21a, A-6020~Innsbruck, Austria}
\affiliation{Parity Quantum Computing GmbH, Rennweg 1, A-6020~Innsbruck, Austria}

\begin{abstract}
The preparation of ground states of spin systems is a fundamental operation in quantum computing and serves as the basis of adiabatic quantum computing. This form of quantum computation is subject to the adiabatic theorem which in turn poses a fundamental speed limit. We show that by employing diabatic transitions via counter diabatic driving a less strict requirement on adiabaticity applies. We demonstrate a scaling advantage from local and multi-spin counter diabatic driving in the ground-state fidelity compared to their adiabatic counterpart, for different Ising spin models.
\end{abstract}

\date{\today}

\maketitle

\section{Introduction}
The recent advances in the field of quantum technology~\cite{buluta2009quantum, cirac2012goals, georgescu2014quantum, saffman2016quantum, preskill2018quantum} have paved the road towards quantum supremacy in the socalled noisy intermediate scale quantum computing (NISQ) regime~\cite{preskill2012quantum, harrow2017quantum}.
Adiabatic quantum computation (AQC) is a computational method for NISQ devices with the aim to solve combinatorial optimization problems~\cite{albash2018adiabatic}.
AQC is subject to the adiabatic theorem~\cite{kato1950adiabatic, messiah1961quantum} which relates the minimal energy gap between the ground state and the first-excited eigenstate and the time to perform an adiabatic evolution. This gap decreases exponentially with the system size at a first-order quantum phase transition (QPT) and polynomially at a second-order QPT~\cite{tsuda2013energy, mishra2018finite}. Thus, the time-to-solution increases exponentially with the size of the system. 
Recent approaches to detach from this fundamental bottleneck have been dubbed \emph{diabatic quantum annealing} (QA)~\cite{crosson2021prospects}. This metaheuristic aims at solving combinatorial optimiziation problems in finite time, i.e. it does not restrict itself to adiabatic driving, thus allowing diabatic transitions between the system eigenstates during the annealing process~\cite{santoro2006optimization, hauke2020perspectives}. The inherent issue with the latter approach, however, is that the success probability to reach the ground state of the system at the end of the QA process decreases considerably, thus leading to a potentially wrong solution to the combinatorial optimization problem of interest.

\par 

A particular form of diabatic quantum annealing is based on counter-diabatic (CD) driving~\cite{arimondo2013chapter, delcampo2015controlling, delcampo2019focus, guery-odelin2019shortcuts}. 
Here, the underlying idea is to suppress unwanted transitions during such finite-time annealing processes by an additional term in the Hamiltonian~\cite{demirplak2003adiabatic, demirplak2005assisted, demirplak2008consistency, berry2009transitionless}.
In general, in order to obtain the exact form of the latter--- that suppresses all transitions between any eigenstates and thus to always find the classical final ground state at the end of the sweep, even for very short durations--- one needs to have a priori knowledge of the system eigenstates at all times during the annealing process, which introduces a severe computational and experimental hindrance for the implementation in near-term quantum devices. To overcome this, a variational approach to find \emph{approximate} CD Hamiltonians has been introduced recently~\cite{sels2017minimizing, claeys2019floquet}. Although approximate CD Hamiltonians are numerically and experimentally much easier to implement, the question how their performance in finding the ground state scales with the system size is still open.

\par 

In this work, we present a polynomial scaling advantage in the ground-state fidelity using approximate CD driving compared to its QA counterpart for Ising spin models with nearest-neighbor and all-to-all connected interactions. For the former, we simulate large system sizes up to $100$ spins using numerical methods based on matrix product states (MPS).
We numerically demonstrate that the exponential complexity of QA reduces to another exponential complexity for approximate CD driving, both with local one-spin and non-local two-spin terms, yet with a considerably smaller coefficient.
We further show for the Ising spin model with all-to-all connectivity, that a similar behavior can be reached, depending on the strength of the interactions between the spins. We further provide a discussion on the nature of the additionally applied CD Hamiltonians and their cost of implementation.
The numerical results reveal a strong polynomial enhancement of the applied approximate CD method, thus serving as an exciting starting point for implementation in near-term quantum annealing devices.

\par 

This work is structured as follows: In Sec.~\ref{sec_methods}, we introduce the methods of approximate CD driving and MPS, present our numerical results in Sec.~\ref{sec_numerical_results} and conclude the latter in Sec.~\ref{sec_discussion}, while giving an outlook on future research and providing supplemental material in the Appendix~\ref{appendix}.

\section{Methods}\label{sec_methods}
Quantum annealing is a metaheuristic to solve combinatorial optimization problems.
In its most general form, it can be written as a time-dependent Hamiltonian of the form 
\begin{align}
\mathcal{H}_0(t) &= [1 - \lambda(t)] \mathcal{H}_\mathrm{d} + \lambda(t) \mathcal{H}_\mathrm{p}, \nonumber \\
\mathcal{H}_\mathrm{d} &= - \sum_{j=1}^N \gamma_j \sigma_j^x
\label{eq:H0}
\end{align}
where $\mathcal{H}_\mathrm{d}$ is the driver Hamiltonian with site-dependent $\sigma^x$, transverse magnetic field strength $\gamma_i$ and $N$ the total number of spins in the system. On the other hand, $\mathcal{H}_\mathrm{p}$ is the problem Hamiltonian in the $z$-direction that encodes our optimiziation problem of interest.
Throughout this work, we use the driving function
\begin{equation}
\lambda(t) = \sin^2\left[\dfrac{\pi}{2}\sin^2\left(\dfrac{\pi t}{2 \tau}\right)\right]
\label{eq_sweep_function}
\end{equation}
that fulfils $\lambda(t=0)=0$ and $\lambda(t=\tau)=1$ with $\tau$ the sweep duration.
The time to find the lowest energy eigenstate, i.e. the ground state, of the problem Hamiltonian $\mathcal{H}_\mathrm{p}$ to solve the optimization problem is subject to the adiabatic theorem~\cite{kato1950adiabatic, messiah1961quantum}. While for long, i.e. adiabatic, sweep durations the annealing schedule $\mathcal{H}_0(t)$, Eq.~\eqref{eq:H0}, the system remains  in the true ground state, the same is not true for shorter times where the success probability drops significantly. 
This dilemma between employing sufficiently short sweep durations while at the same time reaching a large ground-state fidelity constitutes a severe issue within this field.

\par 

To overcome this bottleneck, we additionally employ a counter-diabatic Hamiltonian to speed up QA.
In particular, we drive the quantum system with the total Hamiltonian
\begin{equation}
\mathcal{H}(t)=\mathcal{H}_0(t)+\mathcal{H}_{\mathrm{CD}}(t)
\label{eq_Hsta}
\end{equation}
where $\mathcal{H}_0(t)$, Eq.~\eqref{eq:H0}, is the original Hamiltonian and $\mathcal{H}_\mathrm{CD}(t)$ the additional CD Hamiltonian that aims to suppress transitions between the system's eigenstates and thus to follow the instantaneous ground state of the system.
This time-dependent CD Hamiltonian reads
\begin{equation}
\mathcal{H}_\mathrm{CD}(t) = \dot{\lambda}(t) \mathcal{A}_\lambda(t)
\label{eq_Hcd}
\end{equation}
where $\mathcal{A}_\lambda(t)$ is the \emph{exact} adiabatic gauge potential (AGP)~\cite{sels2017minimizing, kolodrubetz2017geometry, claeys2019floquet} and $\dot{\lambda}(t)$ is the derivative of the sweep function $\lambda(t)$ of Eq.~\eqref{eq_sweep_function}. 
In this work, we employ \emph{approximate} AGPs $\mathcal{A}^\prime_\lambda$ that are much easier to build numerically and experimentally, yet do not fully prevent the system from inducing transitions between the eigenstates during the procedure. 
The corresponding approximate CD Hamiltonians entail $p$-spin terms with an odd number of $\sigma^y$ terms and $p \leq N$. As an example, for $p=1$ we only introduce site-dependent one-spin $\sigma_j^y$ terms, whereas for $p=2$, we additionally employ two-spin $\sigma_j^y \sigma_k^x$ and $\sigma_j^y \sigma_k^z$ terms. 
Employing all combinations of up to $N$-spin terms, i.e. $p=N$, provides us with the \emph{exact} AGP.
In order to obtain the optimal form of these approximate AGPs, we apply an ansatz and minimize the operator distance between the exact AGP and the parameters of the ansatz (see Appendix~\ref{appendix} for more details).

\par 

This method can be applied to various problem Hamiltonians $\mathcal{H}_\mathrm{p}$. Here, we consider two different Ising spin models with (i)~nearest-neighbor and (ii)~all-to-all connected interactions in the problem Hamiltonian $\mathcal{H}_\mathrm{p}$, respectively.\\
The problem Hamiltonian of the Ising spin model with nearest-neighbor interactions and $N$ total spins reads
\begin{align}
\mathcal{H}_\mathrm{p}^{\mathrm{NN}} = - \sum_{j=1}^N b_j \sigma_j^z - \sum_{j=1}^{N-1} J_{j} \sigma_j^z \sigma_{j+1}^z 
\label{eq_Hp_nearest-neighbor}
\end{align}
where $\sigma_j^z$ is the $z$-Pauli matrix at site $j$, $b_j$ is the applied site-dependent longitudinal magnetic field strength and $J_j$ is the interaction strength for which we impose open boundaries, i.e. $\sigma^{N+1}=0$. 
This Ising spin model is described by a one-dimensional short ranged Hamiltonian and thus we can employ numerical methods based on matrix product states (MPS)~\cite{verstraete2006matrix, perez2007matrix, verstraete2008matrix, schollwck2011density, orus2014practical}. In particular, we use a time evolving block decimation (TEBD)~\cite{vidal2004efficient, daley2004time} with a second order Trotter expansion of the Hamiltonian to simulate the dynamics in the MPS.

\par 

As a second Ising spin model, we consider non-local interactions with all-to-all connectivity. The corresponding problem Hamiltonian of interest reads as
\begin{align}
\mathcal{H}_\mathrm{p}^{\mathrm{all}} = - \sum_{j=1}^N b_j \sigma_j^z - \sum_{j=1}^N \sum_{k<j} J_{jk} \sigma_j^z \sigma_{k}^z 
\label{eq_Hp_all-to-all}
\end{align}
where $b_j$ is again the longitudinal magnetic field strength at site $j$ and $J_{jk}$ is the interaction strength between spins at sites $j$ and $k$. Due to the long-range interactions $J_{jk}$, methods based on MPS can not be applied efficiently which restricts our numerical simulations to small system sizes.

\section{Numerical results}\label{sec_numerical_results}
In this section we highlight the performance of finite-time quantum annealing, written in the form of an original Hamiltonian $\mathcal{H}_0(t)$, Eq.~\eqref{eq:H0}, and various counter-diabatic protocols $\mathcal{H}(t) = \mathcal{H}_0(t) + \mathcal{H}_\mathrm{CD}(t)$ with both local one-spin ($p=1$) and non-local two-spin ($p=2$) CD Hamiltonians as in Eq.~\eqref{eq_Hcd}, for different system sizes. 
To this end, we compute for each protocol the success probability, i.e. the final ground-state fidelity 
\begin{equation}
    F(\tau) = |\langle \psi(\tau) | \phi_0(\tau)\rangle|^2
    \label{eq_fidelity}
\end{equation}
as the absolute value of the overlap of the reached state $|\psi(\tau)\rangle$ and $|\phi_0(\tau) \rangle$ the actual ground state at the end of the sweep with duration $\tau$. For the Ising spin chain, we use imaginary-TEBD to compute a MPS representation of the ground state $|\phi_0(\tau) \rangle$.
    
\subsection{Nearest-neighbor model}\label{Sec_NN}
First, we will start with the nearest-neighbor model with original Hamiltonian $\mathcal{H}^\mathrm{NN}_0(t)$, Eq.~\eqref{eq_Hp_nearest-neighbor}. 
The sweep duration is set to $\tau=10$ and the interaction strength to $J_j=0.5$, thus describing a ferromagnet. 
To simulate the dynamics of the system we use a TEBD algorithm with a Trotter time step of $\Delta t = 0.05$ and a maximum bond dimension $\chi=100$. For all the considered systems sizes ($N\leq100$), we verified that the results do not depend on neither $\chi$ nor $\Delta t$.

\par 

\begin{figure}[t]
  \centering
  \includegraphics[width=.95\columnwidth]{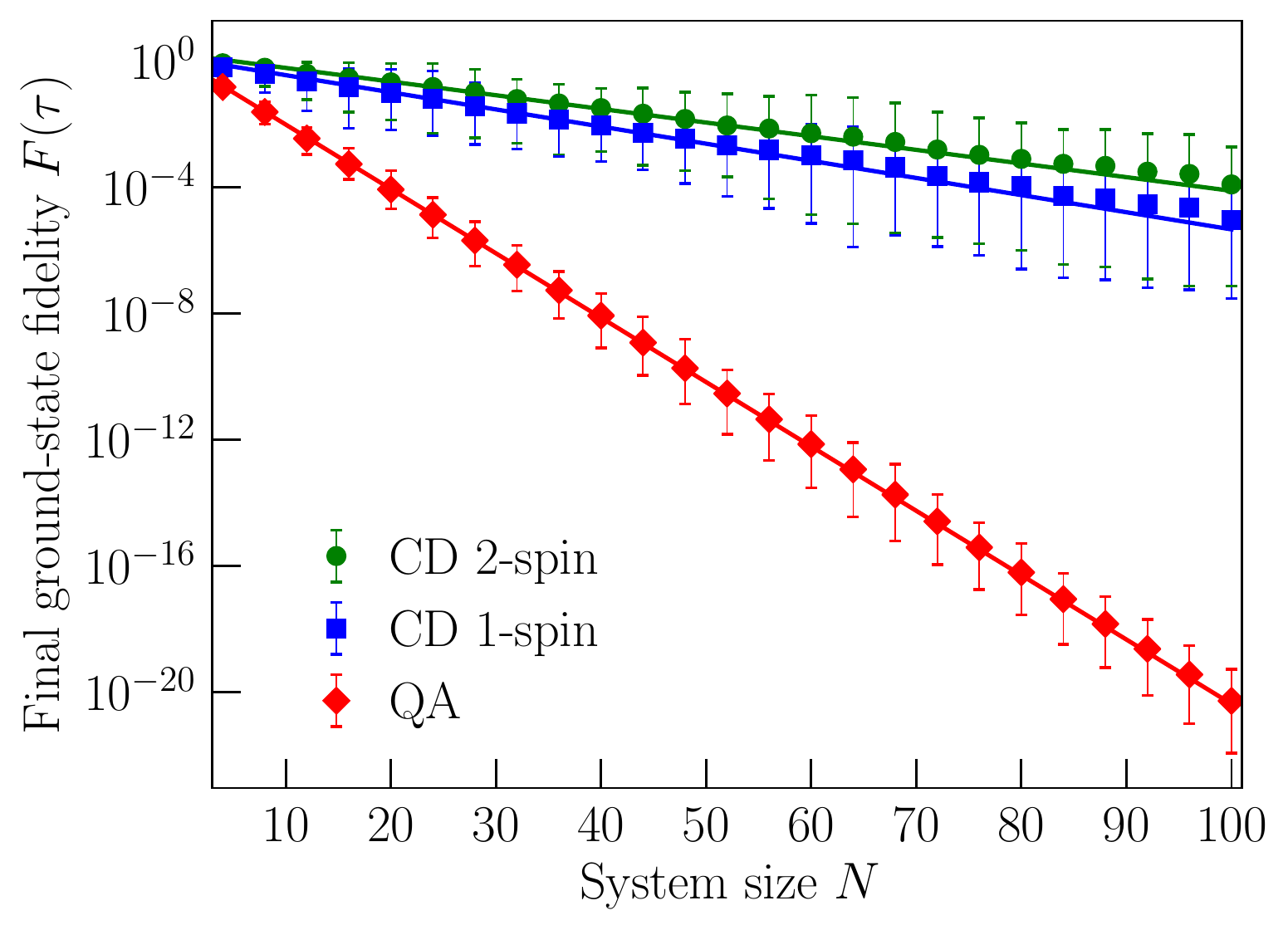}
  \caption{\textbf{Final ground-state fidelity over system size, nearest-neighbor model.} Mean final ground-state fidelity $F(\tau)$, Eq.~\eqref{eq_fidelity}, over the system size $N$ for the nearest-neighbor Ising spin model with (i)~original Hamiltonian [$\mathcal{H}^\mathrm{NN}_0(t)$, Eq.~\eqref{eq_Hp_nearest-neighbor}, red diamonds], (ii)~full Hamiltonian with local one-spin CD [$\mathcal{H}^{(1)}_{\mathrm{CD}}$, Eq.~\eqref{eq_Hcd}, blue squares] and (iii)~non-local two-spin CD protocol [$\mathcal{H}^{(2)}_{\mathrm{CD}}$, Eq.~\eqref{eq_Hcd}, green circles], taken over an ensemble of 50 randomly chosen instances of longitudinal magnetic fields $b_j$, drawn from a Gaussian distribution with mean zero and standard deviation one. The corresponding solid lines depict the fitted exponential curves for each driving protocol with values $a \approx 0.47$ for QA, $b \approx 0.13$ for one-spin and $c \approx 0.1$ for two-spin CD driving, respectively.} 
  \label{fig:fidelity_NN}
\end{figure}

Figure~\ref{fig:fidelity_NN} depicts the reached final ground-state fidelity $F(\tau)$, Eq.~\eqref{eq_fidelity}, for (i)~original [$\mathcal{H}^\mathrm{NN}_0(t)$, Eq.~\eqref{eq_Hp_nearest-neighbor}, red diamonds], (ii)~full Hamiltonian with local one-spin CD protocol [$\mathcal{H}^{(1)}_{\mathrm{CD}}(t)$, Eq.~\eqref{eq_Hcd}, blue squares] and (iii)~non-local two-spin CD protocol [$\mathcal{H}^{(2)}_{\mathrm{CD}}(t)$, Eq.~\eqref{eq_Hcd}, green circles] for an ensemble of 50 randomly chosen longitudinal magnetic fields $b_j$, drawn from a Gaussian distribution with mean zero and standard deviation one, for various system sizes $N$ up to 100 spins. 
For intermediate-time sweeps, the mean of the final ground-state fidelities for QA  with system size $N=100$ drops to  $F_\mathrm{QA}(\tau) \approx 10^{-20}$, whereas for CD driving remains at  $F_\mathrm{CD}^{(1)} \approx 10^{-4}$ for local one-spin and $F_\mathrm{CD}^{(2)} \approx 10^{-3}$ for non-local two-spin CD driving, respectively.

The numerical results reveal a significant enhancement in the ground-state fidelity for the local one-spin and non-local two-spin CD-driven protocols compared to their  quantum annealing counterpart. In particular, the results show a scaling advantage in the sense that the original exponential size complexity $F_0(\tau) \propto \exp(-a N)$ reduces to another exponential complexity, yet with a smaller coefficient, i.e. $F^{(1)}_\mathrm{CD}(\tau) \propto \exp(-b N)$ with $b < a$ for local one-spin CD driving and $F^{(2)}_\mathrm{CD}(\tau) \propto \exp(-c N)$ with $c < b < a$ for non-local two-spin CD driving. 
In particular, we have fitted these curves, i.e. the exponent $s$ of the function $F(N) = r \exp{-s N}$, and received the values $a \approx 0.47$ for QA, $b \approx 0.13$ for one-spin and $c \approx 0.1$ for two-spin CD driving which are plotted as solid lines in figure~\ref{fig:fidelity_NN}.
This scaling advantage in final ground-state fidelity for local and non-local CD driving over quantum annealing persists for various coupling regimes ($J_j = 0.1$ for weak-coupling, $J_j = 0.5$ for intermediate-coupling and $J_j=1$ for the strong-coupling regime, see Appendix~\ref{appendix}).

\subsection{All-to-all connected model}\label{Sec_all}
\begin{figure}[t]
  \centering
  \includegraphics[width=.95\columnwidth]{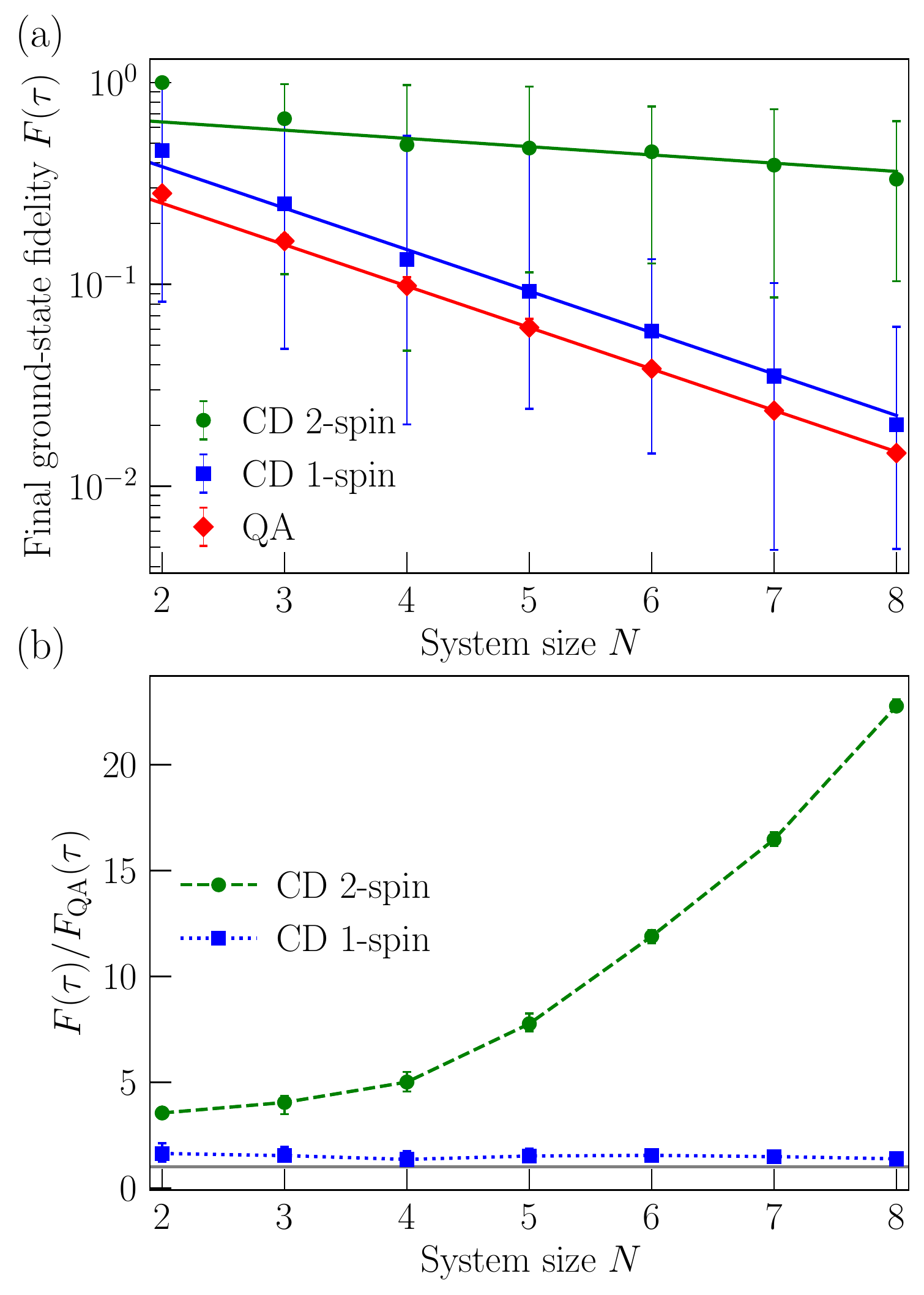}
  \caption{\textbf{Final ground-state fidelity over system size, all-to-all model.} (a)~mean final ground-state fidelity $F(\tau)$, Eq.~\eqref{eq_fidelity}, over the system size $N$ for the all-to-all connected Ising spin model for (i)~original Hamiltonian [$\mathcal{H}^\mathrm{all}_0(t)$, Eq.~\eqref{eq_Hp_all-to-all}, red diamonds], (ii)~full Hamiltonian with local one-spin CD protocol [$\mathcal{H}^{(1)}_{\mathrm{CD}}$, Eq.~\eqref{eq_Hcd}, blue squares] and (iii)~non-local two-spin CD protocol [$\mathcal{H}^{(2)}_{\mathrm{CD}}$, Eq.~\eqref{eq_Hcd}, green circles], taken over an ensemble of 50 randomly chosen instances of longitudinal magnetic fields $b_j$, drawn from a Gaussian distribution with mean zero and standard deviation one. The solid lines depict the fitted exponential curves for each driving protocols with values $a \approx 0.47$ for QA, $b \approx 0.47$ for one-spin and $c \approx 0.1$ for two-spin CD driving, respectively. (b)~quotient of the reached ground-state fidelity for CD and QA protocol, i.e. $F(\tau) / F_\mathrm{QA}(\tau)$. The interaction strength between all combinations of sites is set to $J_{jk}=1$ and the sweep duration to $\tau=1$. The error bars denote largest and lowest value, respectively. The dotted and dashed lines, respectively, depict a mere guide for the eye without fitting.}
  \label{fig:fidelity_all}
\end{figure}

 Let us now consider the Ising spin model with all-to-all connected interactions $J_{jk}$ and original Hamiltonian $\mathcal{H}^\mathrm{all}_0(t)$, Eq.~\eqref{eq_Hp_all-to-all} with transverse magnetic field strengths as well as the interactions strengths between each combination of sites $j$ and $k$ with $\gamma_j = J_{jk}=1$. The sweep duration is $\tau = 1$.
Figure~\ref{fig:fidelity_all}\textcolor{red}{(a)} depicts the reached final ground-state fidelity $F(\tau)$, Eq.~\eqref{eq_fidelity}, for (i)~original [$\mathcal{H}^\mathrm{all}_0(t)$, Eq.~\eqref{eq_Hp_all-to-all}, red diamonds] and full Hamiltonians with (ii)~local one-spin CD [$\mathcal{H}^{(1)}_{\mathrm{CD}}$, Eq.~\eqref{eq_Hcd}, blue squares] and (iii)~non-local two-spin CD protocol [$\mathcal{H}^{(2)}_{\mathrm{CD}}$, Eq.~\eqref{eq_Hcd}, green circles] for an ensemble of 50 randomly chosen longitudinal magnetic fields $b_j$, drawn from a Gaussian distribution with mean zero and standard deviation one, for various system sizes $N$ up to eight spins.
The mean of the final ground-state fidelities for quantum annealing approximately scale as $F_\mathrm{QA}(\tau) \propto 1 / 2^N$. This is the parameter regime of the quench limit where the spins randomly flip from the $x$- into the $z$-direction, each pointing with probability $1/2$ into the right direction. Local CD driving provides a constant speedup compared to QA. On the other hand, the fidelities with non-local two-spin CD driving depict a polynomial scaling enhancement for increasing system sizes. For this model, the fitted exponents in the exponential curves read $a \approx 0.47 \approx b$ for QA and local one-spin CD driving, respectively, and $c \approx 0.1$ for non-local two-spin CD driving.\\
Figure~\ref{fig:fidelity_all}\textcolor{red}{(b)} depicts the quotient of the reached final ground-state fidelity for CD and traditional QA driving, i.e. $F(\tau) / F_\mathrm{QA}(\tau)$ where for convenience we provided a guide for the eye. In this panel, we see that a scaling advantage for non-local two-spin CD driving in the reached final ground-state fidelities can be reached towards their traditional QA counterpart.

\par 

The numerical results reveal a considerable enhancement in the ground-state fidelity, similar to systems with nearest-neighbor interaction. Whereas for the latter the polynomial enhancement persists also for local one-spin CD driving, this feature ceases to exist for interactions with all-to-all connectivity due to its  increased complexity and seem to diminish to a constant speedup only. However, for non-local two-spin CD driving the polynomial enhancement is observed as approximate AGP also contains two-spin terms to account for the dominance of the all-to-all connected interactions.

\section{Discussion and outlook}\label{sec_discussion}
In this work, we have numerically demonstrated a scaling advantage in the  ground-state fidelity using  approximate local and non-local counter-diabatic driving compared to their quantum annealing counterparts, for two different classes of Ising spin models. The enhancement is strongly polynomial, meaning that the original exponential size complexity remains exponential for approximate CD driving, yet with a smaller coefficient. Remarkably, this feature also holds true for local CD driving whose full Hamiltonian can, by rotation around the z-axis, be brought into the form of a transverse field Ising model, consisting of merely $\sigma^x$ and $\sigma^z$ terms, yet with unconventional driving of the system parameters. 
We expect the scaling advantage to increase with larger $p$-spin CD driving, potentially up to an ideal exponential speedup for $N$-spin CD driving, where we adapt the \emph{exact} adiabatic gauge potential, thus always tracking the instantaneous ground state for all sweep durations.
This feature comes with an emerging cost of implementing the multi-spin CD protocols numerically as well as experimentally. 
The algebraic or numerical determination of the optimal parameters of the corresponding CD Hamiltonians and AGPs suffers from a polynomial overhead for increasing $p$-spin CD driving. On the other hand, implementing non-local CD Hamiltonians experimentally is hard to accomplish for current quantum annealing devices. In the end, these numerical results reveal a trade-off between a further enhancement of performance and the cost of implementation of the additional non-local AGPs.
The existing polynomial scaling advantage for the case of local CD-driving compared to traditional QA encourage the experimental implementation of the former and may serve as an excellent candidate for efficient devices by using further modifications of these methods (cf. Refs.~\cite{hartmann2019rapid, prielinger2021twoparameter}) which is an interesting future topic to be considered.

\begin{acknowledgements}
This work was supported by the Austrian Science Fund (FWF) through a START grant under Project No. Y1067-N27 and the SFB BeyondC Project No. F7108-N38, the Hauser-Raspe foundation, and the European Union's Horizon 2020 research and innovation program under grant agreement No. 817482. This material is based upon work supported by the Defense Advanced Research Projects Agency (DARPA) under Contract No. HR001120C0068. Any opinions, findings and conclusions or recommendations expressed in this material are those of the author(s) and do not necessarily reflect the views of DARPA.
\end{acknowledgements}

\appendix

\section{Approximate counter-diabatic driving}\label{appendix}
In order to find the optimal form of the adiabatic gauge potential $\mathcal{A}_\lambda(t)$ for the additional counter-diabatic Hamiltonian $\mathcal{H}_\mathrm{CD}(t)$ from Eq.~\eqref{eq_Hcd}, we employ the method introduced in Refs.~\cite{sels2017minimizing, kolodrubetz2017geometry}. In this approach, we rely on an \emph{approximate} AGP $\mathcal{A}^\prime_\lambda(t)$ that can be derived from a variational principle. To this end, we define a Hermitian operator of interest as
\begin{equation}
    G_\lambda(\mathcal{A}^\prime_\lambda) = \partial_\lambda \mathcal{H}_0 + i [\mathcal{A}^\prime_\lambda, \mathcal{H}_0]
    \label{App_eq_hermitian_operator}
\end{equation}
where we have set $\hbar \equiv 1$.
With this, we can further define an operator distance between the exact solution for the AGP, $\mathcal{A}_\lambda$, and our ansatz for the approximate solution, $\mathcal{A}^\prime_\lambda$, as the Frobenius norm
\begin{align}
\mathcal{D}(\mathcal{A}^\prime_\lambda) = \mathrm{Tr}[ (G_{\lambda}(\mathcal{A}'_\lambda) + \mathcal{M}_{\lambda})^2] = \mathrm{Tr}[G^2_\lambda(\mathcal{A}'_\lambda)]  - \mathrm{Tr}[\mathcal{M}_{\lambda}^2]
	\label{App_A_eq_operatordistance}
\end{align}
where $\mathcal{M}_\lambda$ describes the generalized force. 
In consequence, we aim to minimize this operator distance to find the optimal form of the approximate AGP $\mathcal{A}^\prime_\lambda$. 
As the second term of Eq.~\eqref{App_A_eq_operatordistance} is independent of the parameters in the ansatz for the approximate AGP, minimization of this operator distance is equivalent to minimizing the action
\begin{equation}
    \mathcal{S}(\mathcal{A}^\prime_\lambda) = \mathrm{Tr}[G^2_\lambda(\mathcal{A}^\prime_\lambda)]
    \label{App_eq_S}
\end{equation}
with respect to the parameters of the ansatz for $\mathcal{A}^\prime_\lambda$, symbolically written as
\begin{equation}
    \dfrac{\delta S(\mathcal{A}^\prime_\lambda)}{\delta \mathcal{A}^\prime_\lambda} = 0
    \label{App_eq_dS}
\end{equation}
where the symbol $\delta$ depicts the functional derivative.

\par 

In this work, we rely on two kinds of approximate CD Hamiltonians $\mathcal{H}_\mathrm{CD}(t)$.
First, a local one where the ansatz for the AGP contains single-spin terms and, second, non-local ans\"atze where the corresponding AGP contains multi-spin operators.

\subsubsection*{Implementation cost} 
The implementation of these additional CD Hamiltonians comes with a price to pay that we consider hereinafter. To estimate these costs, various quantifiers have been developed recently~\cite{zheng2016cost, campbell2017tradeoff}. In this work, we employ the quantifier
\begin{equation}
\langle \mathcal{H}_\mathrm{CD} \rangle = \nu_{t,N} \int_0^\tau \mathrm{Tr}[\mathcal{H}^\dagger_\mathrm{CD}(t) \mathcal{H}_\mathrm{CD}(t)] dt
\label{App_eq_implementation_cost}
\end{equation}
that determines the power that is needed to generate the additional magnetic fields due to the presence of these CD Hamiltonians. Here, $\nu_{t,N}$ is a highly setup-dependent parameter that in general is hard to be determined and which is set to one during the following additional numerical results.

\subsection{Nearest-neighbor model}\label{app_NN}
For the Ising spin model with nearest-neighbor interactions $J_j$, we employ the original Hamiltonian
\begin{multline}
    \mathcal{H}_0^\mathrm{NN}(t)= - [1 - \lambda(t)] \sum_{j=1}^N \gamma_j \sigma_j^x \\ - \lambda(t) \left[ \sum_{j=1}^N b_j \sigma_j^z + \sum_{j=1}^{N-1} J_j \sigma_j^z \sigma_{j+1}^z \right] 
    \label{App_eq_H0_NN}
\end{multline}
where the driving function $\lambda(t)$ is chosen according to Eq.~\eqref{eq_sweep_function}, fulfilling the conditions $\lambda(t=0)=0$ and $\lambda(t=\tau)=1$ with $\tau$ the sweep duration, $\gamma_j$ as well as $b_j$ the site-dependent transverse field and longitudinal magnetic field strengths, respectively, and $J_j$ the interaction strength between spins at sites $j$ and $j+1$, for which we apply open boundary conditions, i.e. $\sigma_{N+1}^z=0$.

\par 

For this model we choose two types of AGPs: (i)~a local single-spin, $\mathcal{A}^\mathrm{(1)}_\lambda$, and (ii)~non-local two-spin, $\mathcal{A}^\mathrm{(2)}_\lambda$, ansatz for the approximate AGPs, i.e.,
\begin{align}
    \mathcal{A}^\mathrm{(1)}_\lambda &= \sum_{j=1}^N \alpha_j \sigma_j^y, \\
    \mathcal{A}^\mathrm{(2)}_\lambda &= \sum_{j=1}^N \alpha_j \sigma_j^y + \sum_{j=1}^{N-1} \beta_j \sigma_j^y \sigma_{j+1}^x + \gamma_j \sigma_j^x \sigma_{j+1}^y \nonumber \\
    &+ \delta_j \sigma_j^y \sigma_{j+1}^z + \epsilon_j \sigma_j^z \sigma_{j+1}^y,
    \label{App_eq_AGP_local_non-local_NN}
\end{align}
where we determine the site-dependent parameters $\alpha_j$, $\beta_j$, $\gamma_j$, $\delta_j$ and $\epsilon_j$ by minimizing the corresponding action~\eqref{App_eq_S} numerically. For this model, the number of parameters to be determined is $N$ for the local, and $N+4(N-1) = 5N -4$ for the non-local ansatz, respectively.

\subsubsection*{Additional numerical results}
\begin{figure}[t]
  \centering
  \includegraphics[width=\columnwidth]{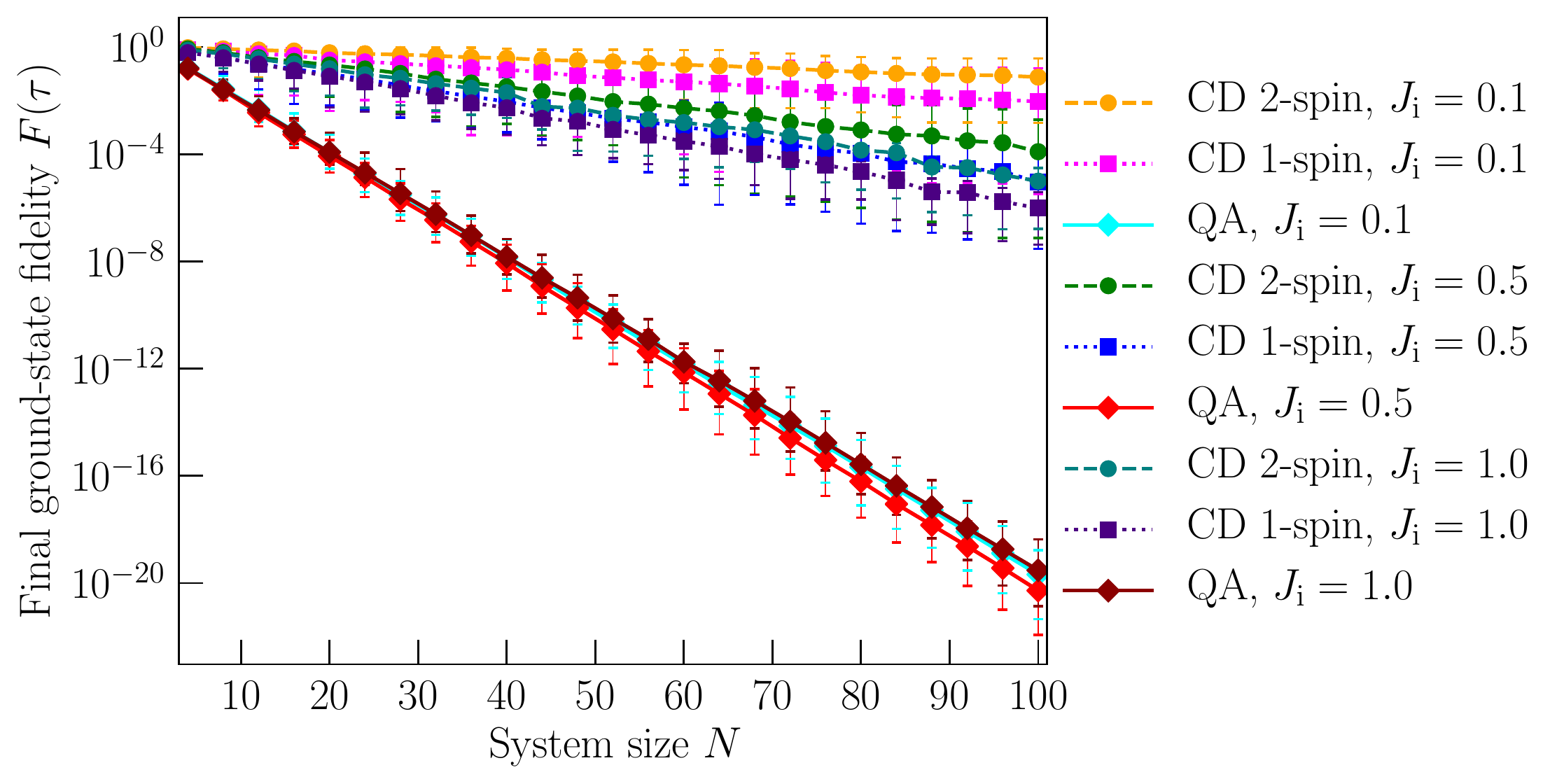}
  \caption{\textbf{Final ground-state fidelity over system size, nearest-neighbor model.} Mean final ground-state fidelity $F(\tau)$, Eq.~\eqref{eq_fidelity}, over the system size $N$ for the nearest-neighbor Ising spin model for (i)~original Hamiltonian [$\mathcal{H}^\mathrm{NN}_0(t)$, Eq.~\eqref{eq_Hp_nearest-neighbor}, solid diamonds], (ii)~full Hamiltonian with local one-spin CD protocol [$\mathcal{H}^{(1)}_{\mathrm{CD}}$, Eq.~\eqref{eq_Hcd}, dotted squares] and (iii)~non-local two-spin CD protocol [$\mathcal{H}^{(2)}_{\mathrm{CD}}$, Eq.~\eqref{eq_Hcd}, dashed circles], taken over an ensemble of 50 randomly chosen instances of longitudinal magnetic field strengths $b_j$, drawn from a Gaussian distribution with mean zero and standard deviation one, for different interaction strengths $J_j$. The sweep duration is set to $\tau=10$ and the transverse magnetic field strengths to $\gamma_j=1$. The error bars denote largest and lowest value, respectively.}
  \label{App:fidelity_NN}
\end{figure}

Figure~\ref{App:fidelity_NN} depicts the reached final-ground state fidelity $F(\tau)$, Eq.~\eqref{eq_fidelity}, over the system size $N$ for the same parameters as in the main text (\cf Fig.~\ref{fig:fidelity_NN} in Sec.~\ref{Sec_NN}), yet for different interaction strengths $J_j$, reaching from weakly ($J_j = 0.1$) via intermediate ($J_j = 0.5$) to strongly coupled spins ($J_j = 1$). 
The numerical results reveal that the polynomial scaling advantage in reached final ground-state fidelity remains stable even for different interaction strengths, in particular, in the strong-coupling regime.

\begin{figure}[ht]
  \centering
  \includegraphics[width=0.8\columnwidth]{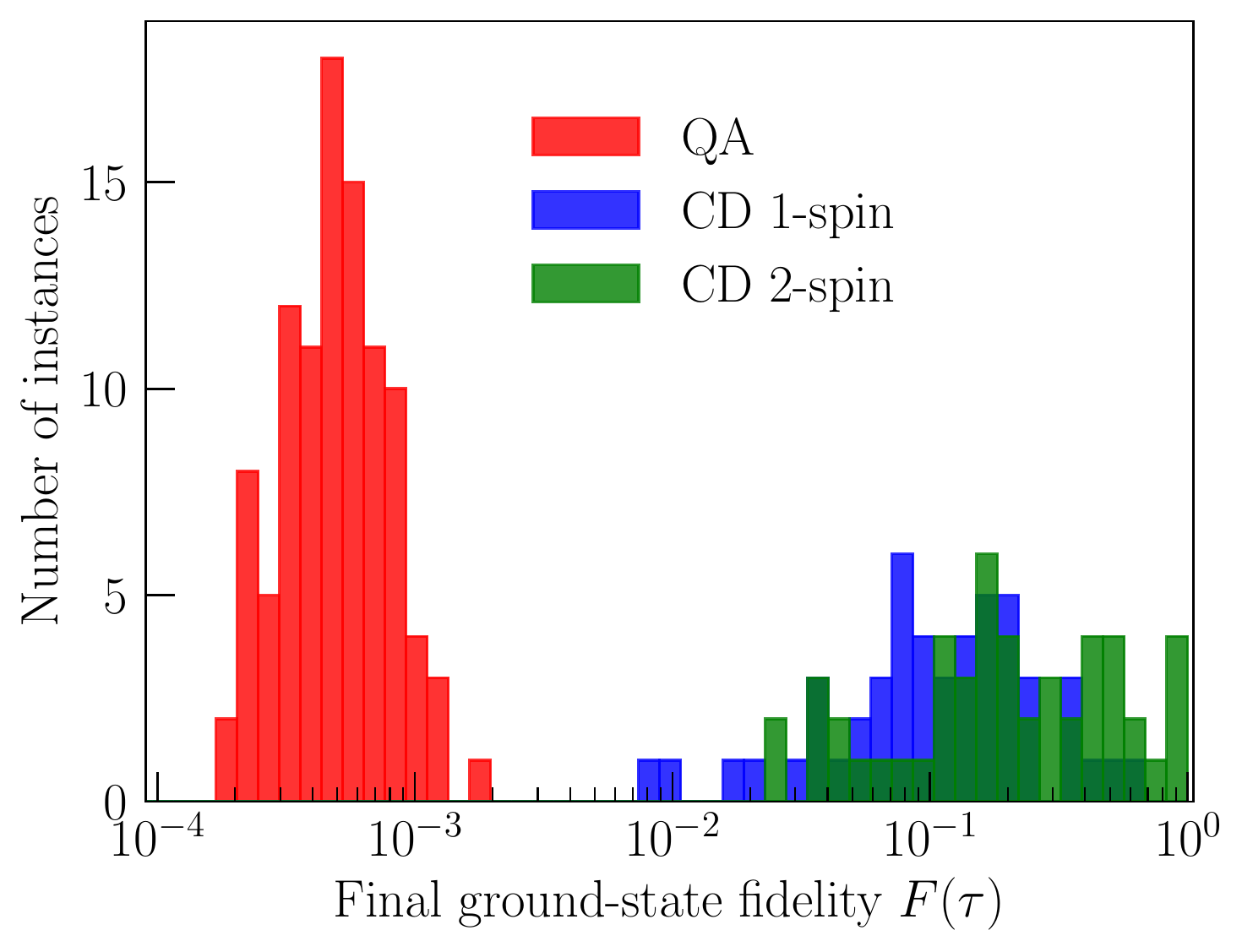}
  \caption{\textbf{Number of instances over ground-state fidelity.} Histogram depicts the number of instances that attain a certain reached final ground-state fidelity $F(\tau)$, Eq.~\eqref{eq_fidelity}, for the nearest-neighbor Ising spin model for (i)~original Hamiltonian [$\mathcal{H}^\mathrm{NN}_0(t)$, Eq.~\eqref{eq_Hp_nearest-neighbor}, red], (ii)~full Hamiltonian with local one-spin CD protocol [$\mathcal{H}^{(1)}_{\mathrm{CD}}$, Eq.~\eqref{eq_Hcd}, blue] and (iii)~non-local two-spin CD protocol [$\mathcal{H}^{(2)}_{\mathrm{CD}}$, Eq.~\eqref{eq_Hcd}, green], taken over an ensemble of 50 randomly chosen instances of longitudinal magnetic fields $b_j$, drawn from a Gaussian distribution with mean zero and standard deviation one, for system size $N=16$ and other parameters as in Fig.~\ref{fig:fidelity_NN}.}
  \label{App:fig_histogramm_NN}
\end{figure}

Furthermore, we are interested in the distribution of the reached final ground-state fidelities.
Figure~\ref{App:fig_histogramm_NN} depicts the latter for system size $N=16$ and the same ensemble of chosen longitudinal magnetic field strengths $b_j$ and other parameters as in Fig.~\ref{fig:fidelity_NN}.
The numerical results depict that the distribution of the reached ground-state fidelities $F(\tau)$ for the single instances can approximately be regarded as a Gaussian function for all three protocols, traditional quantum annealing (red bars), local one-spin CD driving (blue) and non-local two-spin CD driving (green). The reached fidelities for the latter two, however, are highly increased compared to their traditional QA counterpart.

\par
\begin{figure}[htbp]
  \centering
  \includegraphics[width=0.8\columnwidth]{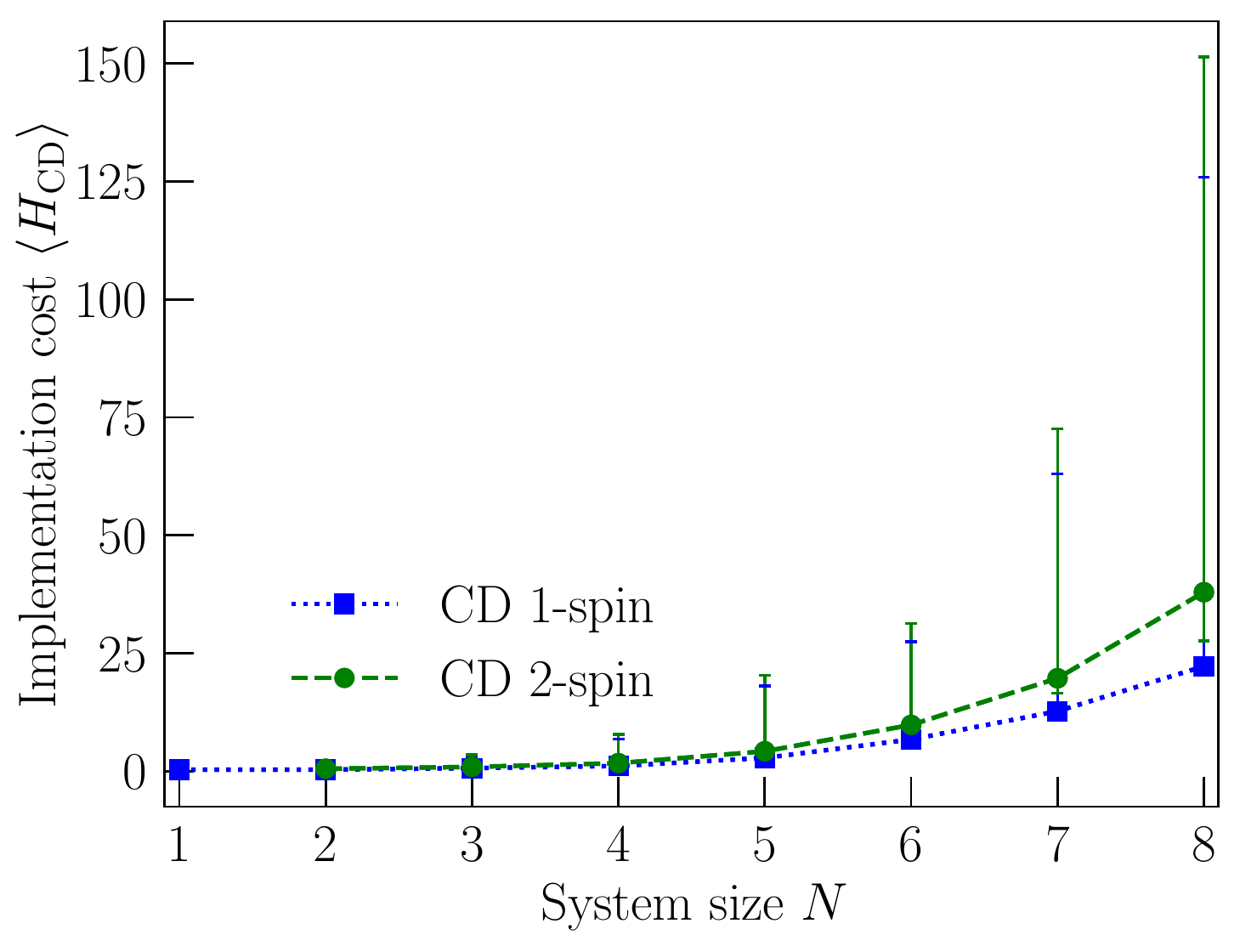}
  \caption{\textbf{Final ground-state fidelity.} Mean final ground-state fidelity $F(\tau)$, Eq.~\eqref{eq_fidelity}, over the system size $N$ for the nearest-neighbor Ising spin model for (i)~full Hamiltonian with local one-spin CD protocol [$\mathcal{H}^{(1)}_{\mathrm{CD}}$, Eq.~\eqref{eq_Hcd}, blue squares] and (ii)~non-local two-spin CD protocol [$\mathcal{H}^{(2)}_{\mathrm{CD}}$, Eq.~\eqref{eq_Hcd}, green circles], taken over an ensemble of 50 randomly chosen instances of longitudinal magnetic fields $b_j$, drawn from a Gaussian distribution with mean zero and standard deviation one. The sweep duration is set to $\tau=10$ the transverse magnetic field strengths to $\gamma_j=1$ and interactions strengths to $J=0.5$. The error bars denote largest and lowest value.}
  \label{App:fig_energetic_cost_NN}
\end{figure}

 
Figure~\ref{App:fig_energetic_cost_NN} depicts the implementation cost [$\langle \mathcal{H}_\mathrm{CD} \rangle$, Eq.~\eqref{App_eq_implementation_cost}] for (i)~local one-spin CD driving (blue squares) and (ii)~non-local two-spin CD driving (green circles) for the same ensemble of instances and various system sizes $N$ as described in Sec.~\ref{Sec_NN}.
We see that these implementation costs scale polynomially with increasing system size $N$ for both local and non-local CD driving. For the latter, the slope is larger as we have to take the additional $4(N-1)$ two-spin terms into account compared to only $N$ one-spin terms from local CD driving.

\subsection{All-to-all connected model}\label{app_all-to-all}
For the Ising spin model with all-to-all connectivity, we employ the original Hamiltonian
\begin{multline}
        \mathcal{H}_0^\mathrm{all}(t)= - [1 - \lambda(t)] \sum_{j=1}^N \gamma_j \sigma_j^x \\ - \lambda(t) \left[ \sum_{j=1}^N b_j \sigma_j^z + \sum_{j=1}^{N-1} \sum_{k > j} J_{jk} \sigma_j^z \sigma_{k}^z \right]
    \label{App_eq_H0_all}
\end{multline}
where in analogy to the nearest-neighbor model we choose the driving function $\lambda(t)$ according to Eq.~\eqref{eq_sweep_function}, and $\gamma_j$ and $b_j$ describe the site-dependent transverse and longitudinal magnetic field strengths, respectively, and $J_{jk}$ the interaction strength between spins at sites $j$ and $k$.

\par 

In analogy to the nearest-neighbor model, we choose a local single-spin, $\mathcal{A}^\mathrm{(1)}_\lambda$, and non-local two-spin, $\mathcal{A}^\mathrm{(2)}_\lambda$, ansatz for the approximate AGPs, i.e.,
\begin{align}
    \mathcal{A}^\mathrm{(1)}_\lambda &= \sum_{j=1}^N \alpha_j \sigma_j^y, \\
    \mathcal{A}^\mathrm{(2)}_\lambda &= \sum_{j=1}^N \alpha_j \sigma_j^y + \sum_{j=1}^{N-1} \sum_{k > j} \beta_{jk} \sigma_j^y \sigma_{k}^x + \gamma_{jk} \sigma_j^x \sigma_{k}^y \nonumber \\
    &+ \delta_{jk} \sigma_j^y \sigma_{k}^z + \epsilon_{jk} \sigma_j^z \sigma_{k}^y,
    \label{App_eq_AGP_local_non-local_NN}
\end{align}
where we determine the site- and interaction-dependent parameters $\alpha_j$, $\beta_{jk}$, $\gamma_{jk}$, $\delta_{jk}$ and $\epsilon_{jk}$ by minimizing the corresponding action~\eqref{App_eq_S} numerically.
For this model with all-to-all connected interactions and non-local ansatz of the corresponding AGP, the number of parameters to be determined is $N + 4N(N-1)/2 = 4N^2 - N$ and thus scales quadratically with the system size instead of linearly as for the local ansatz and thus may constitute a numerical bottleneck for larger system sizes $N$. 

\subsubsection*{Additional numerical results}
\begin{figure}[t]
  \centering
  \includegraphics[width=0.8\columnwidth]{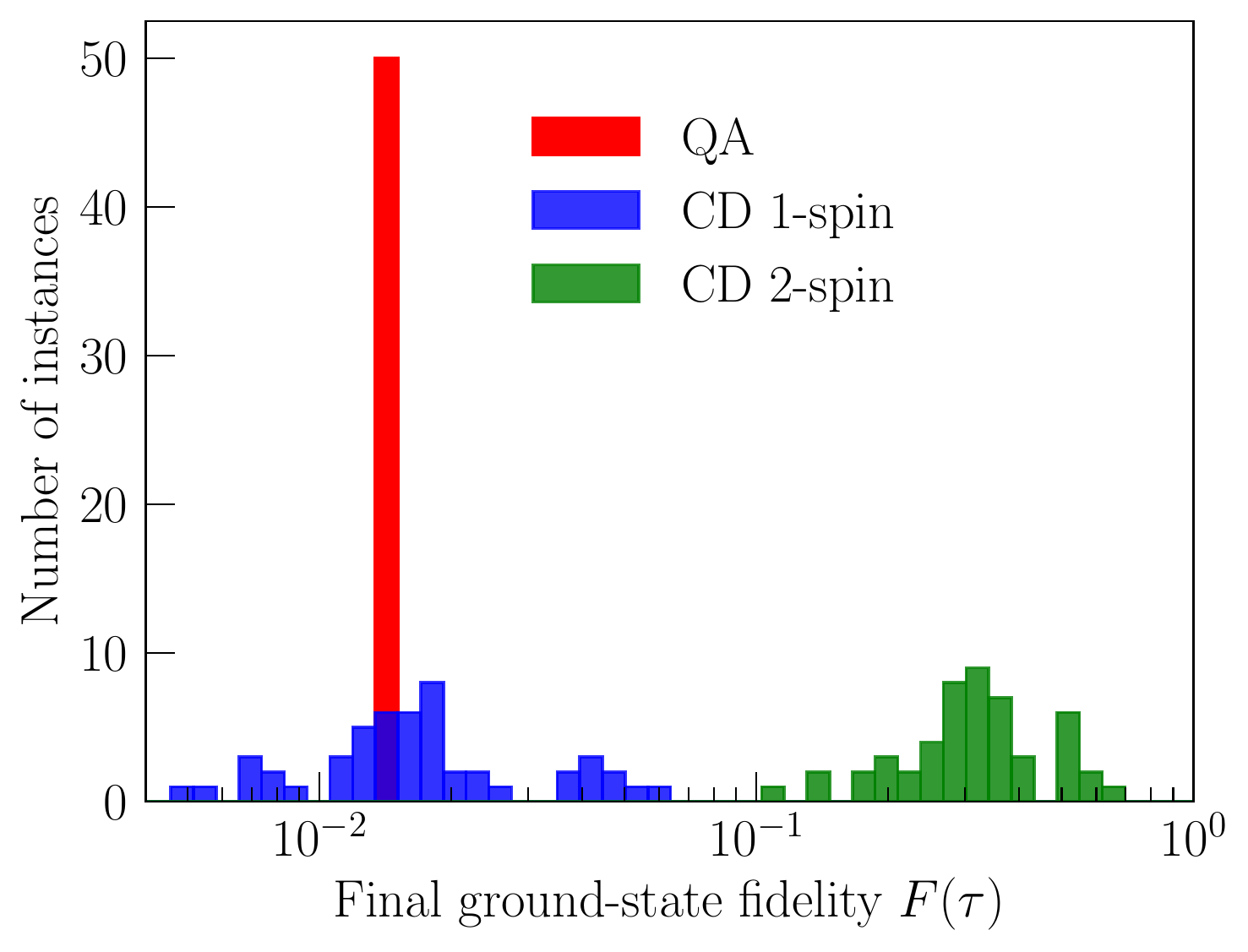}
  \caption{\textbf{Number of instances over ground-state fidelity.} Number of instances that attain a certain reached final ground-state fidelity $F(\tau)$, Eq.~\eqref{eq_fidelity}, for the all-to-all connected Ising spin model for (i)~original Hamiltonian [$\mathcal{H}^\mathrm{all}_0(t)$, Eq.~\eqref{eq_Hp_all-to-all}, red], (ii)~full Hamiltonian with local one-spin CD protocol [$\mathcal{H}^{(1)}_{\mathrm{CD}}$, Eq.~\eqref{eq_Hcd}, blue] and (iii)~non-local two-spin CD protocol [$\mathcal{H}^{(2)}_{\mathrm{CD}}$, Eq.~\eqref{eq_Hcd}, green], taken over an ensemble of 50 randomly chosen instances of longitudinal magnetic fields $b_j$, drawn from a Gaussian distribution with mean zero and standard deviation one, for system size $N=8$ and other parameters as in Fig.~\ref{fig:fidelity_all}.}
  \label{App:fig_histogramm_all}
\end{figure}
Figure~\ref{App:fig_histogramm_all} depicts the distribution of the reached final ground-state fidelities $F(\tau)$ of Eq.~\eqref{eq_fidelity} for a system size $N=8$ and the same ensemble of chosen longitudinal magnetic field strengths $b_j$ and other parameters as in Fig.~\ref{fig:fidelity_all}.
We see that the reached final ground-state fidelities for traditional quantum annealing (red bars) are all approximately equal, thus presenting the case of the quench limit where the fidelities scale as $F(\tau) \propto 1 / 2^N$. For local one-spin (blue) and non-local two-spin CD driving (green) the reached final fidelities are both Gaussian distributed. Whereas the fidelities for one-spin CD driving are only slightly larger than for the traditional quantum annealing counterpart while some instances perform even worse, all instances for the case of two-spin CD driving are considerably enhanced.

\par

\begin{figure}[htpb]
  \centering
  \includegraphics[width=0.8\columnwidth]{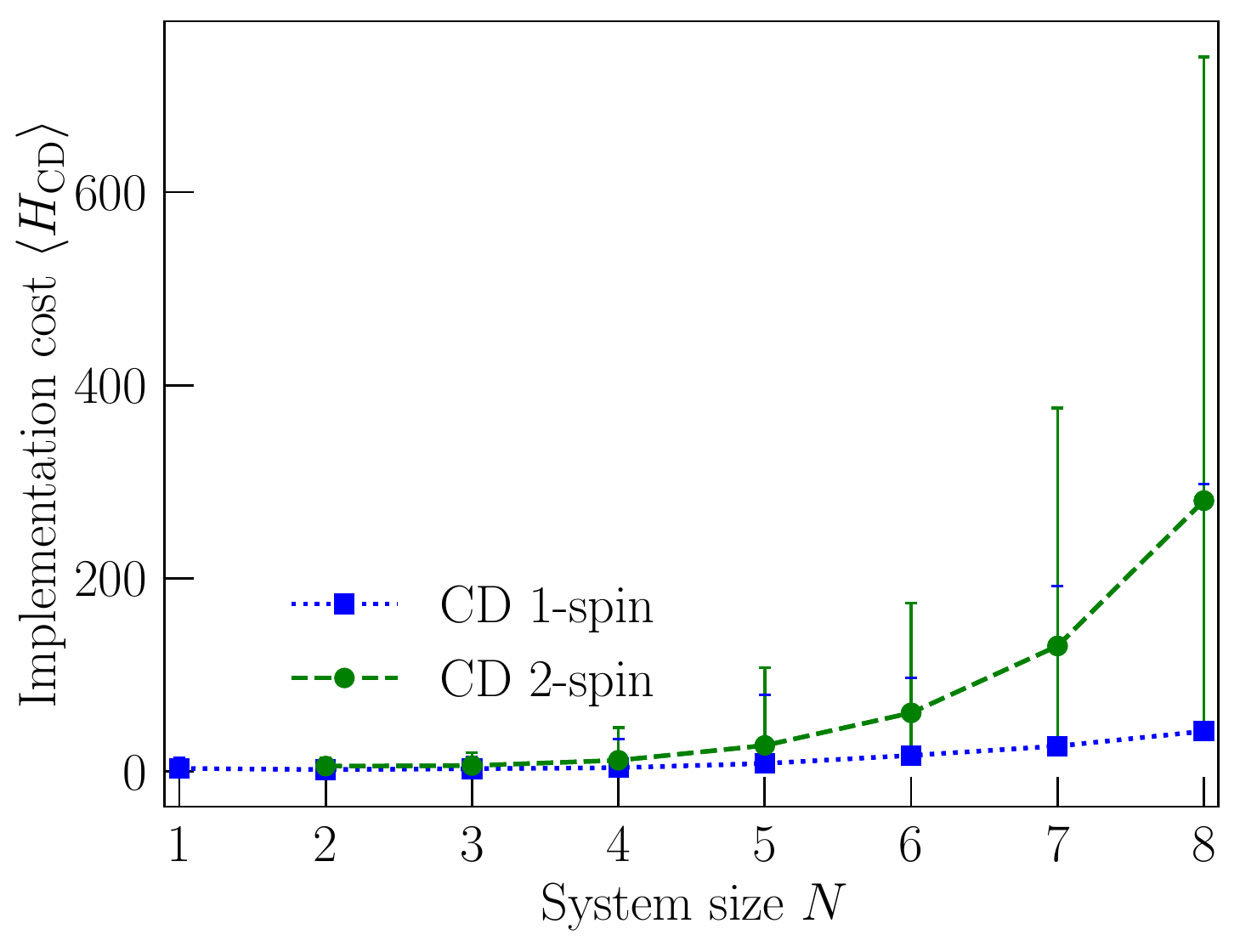}
  \caption{\textbf{Final ground-state fidelity.} Mean final ground-state fidelity $F(\tau)$, Eq.~\eqref{eq_fidelity}, over the system size $N$ for the all-to-all connected Ising spin model for (i)~full Hamiltonian with local one-spin CD protocol [$\mathcal{H}^{(1)}_{\mathrm{CD}}$, Eq.~\eqref{eq_Hcd}, blue squares] and (ii)~non-local two-spin CD protocol [$\mathcal{H}^{(2)}_{\mathrm{CD}}$, Eq.~\eqref{eq_Hcd}, green circles], taken over an ensemble of 50 randomly chosen instances of longitudinal magnetic fields $b_j$, drawn from a Gaussian distribution with mean zero and standard deviation one. The sweep duration is set to $\tau=10$ and the transverse magnetic field and interactions strengths to $\gamma_j= J_{jk} = 1$. The error bars denote largest and lowest value.}
  \label{App_energetic_cost_all}
\end{figure}
Figure~\ref{App_energetic_cost_all} depicts the implementation cost $\langle \mathcal{H}_\mathrm{CD} \rangle$, Eq.~\eqref{App_eq_implementation_cost}, for the same ensembles and system sizes as described in Sec.~\ref{Sec_all}, thus for (i)~local CD driving (blue squares) and (ii)~non-local CD driving (green circles).
The numerical results show that these costs also increase polynomially with the system size, yet the slope of the non-local two-spin curve is highly enlarged as -- in contrast to the nearest-neighbor model -- the number of parameters to be determined for the CD Hamiltonian now scales as $4N^2 - N$ and thus has a quadratic overhead compared to applying local terms only or the non-local terms in the nearest-neighbor model.

\end{document}